\documentclass[acmtog]{acmart}

\usepackage{booktabs} 

\usepackage[ruled]{algorithm2e} 

\SetAlFnt{\small}
\SetAlCapFnt{\small}
\SetAlCapNameFnt{\small}
\SetAlCapHSkip{0pt}
\IncMargin{-\parindent}

\setcopyright{none}

\acmPrice{0}

\usepackage{url}
\usepackage{multirow}
\usepackage{tabularx}
\usepackage{subfig}

\newcommand{\fps}{~\text{fps}}
\newcommand{\by}{{\times}}

\begin{document}

\title{Rendering of Complex Heterogenous Scenes using Progressive Blue Surfels}

\author{Sascha Brandt}
\affiliation{%
  \department{Heinz Nixdorf Institute}
  \institution{Paderborn University}
  \streetaddress{F\"urstenallee 11}
  \city{Paderborn}
  \country{Germany}
}
\email{sascha.brandt@upb.de}

\author{Claudius J\"ahn}%
\affiliation{%
  \institution{DeepL.com}
  \streetaddress{Im Mediapark 8a}
  \city{Cologne}
  \country{Germany}
}
\email{claudius@deepl.com}

\author{Matthias Fischer}
\affiliation{%
  \department{Heinz Nixdorf Institute}
  \institution{Paderborn University}
  \streetaddress{F\"urstenallee 11}
  \city{Paderborn}
  \country{Germany}
}
\email{mafi@upb.de}

\author{Friedhelm Meyer auf der Heide}
\affiliation{%
  \department{Heinz Nixdorf Institute}
  \institution{Paderborn University}
  \streetaddress{F\"urstenallee 11}
  \city{Paderborn}
  \country{Germany}
}
\email{fmadh@upb.de}

\renewcommand\shortauthors{Brandt, S. et al}

\begin{abstract}
We present a technique for rendering highly complex 3D scenes in real-time by generating uniformly distributed points on the scene's visible surfaces.
The technique is applicable to a wide range of scene types, like scenes directly based on complex and detailed CAD data consisting of billions of polygons (in contrast to scenes handcrafted solely for visualization).
This allows to visualize such scenes smoothly even in VR on a HMD with good image quality, while maintaining the necessary frame-rates.
In contrast to other point based rendering methods, we place points in an approximated blue noise distribution only on visible surfaces and store them in a highly GPU efficient data structure, allowing to progressively refine the number of rendered points to maximize the image quality for a given target frame rate.
Our evaluation shows that scenes consisting of a high amount of polygons can be rendered with interactive frame rates with good visual quality on standard hardware.
\end{abstract}

\ccsdesc[500]{Computing methodologies~Rasterization}
\ccsdesc[300]{Computing methodologies~Visibility}
\ccsdesc[300]{Computing methodologies~Image-based rendering}
\ccsdesc[300]{Computing methodologies~Virtual reality}
\ccsdesc[100]{Computing methodologies~Point-based models}

\keywords{surfels, splats, point rendering, continuous level of detail, blue noise, virtual reality, CAD}

\begin{teaserfigure}
  \includegraphics[width=\textwidth]{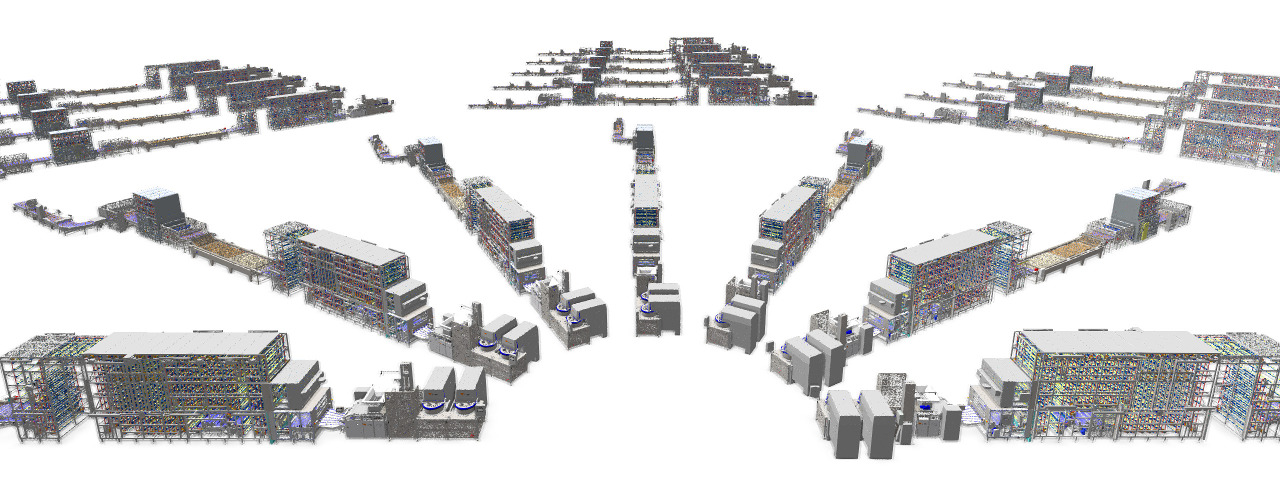}
  \caption{20 donut production lines with ${\sim}50M$ Polygons each (${\sim}1G$ total) rendered at interactive frame rates of ${\sim}120\fps$ at a resolution of $2560{\times}1440$.}
  \label{fig:teaser}
\end{teaserfigure}


\maketitle
\thispagestyle{empty}

\section{Introduction}
\label{sec:intro}

Real time rendering of highly complex scenes used to be a major topic in computer graphics for a considerable time.
In recent years, the interest dropped as one might consider the problem solved considering the vast computational power of modern graphics hardware and the rich set on different rendering techniques like culling and approximation techniques.

A new challenge arises from the newest generation of head mounted displays (HMD) like the Oculus Rift or its competitors -- high framerates are no longer a nice-to-have feature but become a hard constraint on what content can actually be shown to the user.
To avoid nausea it is necessary to keep a high frame-rate of $90fps$ while rendering in stereo at a high resolution.
Although, the lower frame-rate limit of $90fps$ can be reduced to $45fps$ by using techniques such as time warping \cite{Waveren2016}, it is still challenging to display highly complex 3d scenes in stereo at the required resolutions.
Furthermore, due to the distortion by the HMD lenses, it is usually required to render to an oversized frame buffer that exceeds the HMDs resolution to achieve the required image quality at the focal area.
But, this also means, that pixels at the border get skewed and we would render unnecessary high details.
Modern game engines adapted to these requirements and allow the rendering of virtual scenes with an impressive visual quality on head mounted displays; but only if the scenes are specially designed for visualization purposes or have a low complexity.

Another challenge is posed by the rendering of complex CAD data.
It can be a requirement to visualize multiple machines in a virtual machine hall for interactive design reviews for, e.g., planning purposes.
These virtual prototypes are often visualized in a big cave-like system with multiple projection screens to provide a multi-user virtual reality view on the 3d scene to analyse and discuss potential problems and improvements of a technical system.
However, the real-time requirements for such interactive virtual design reviews are usually much lower than for, e.g., games ($20-30fps$) and for the image quality, visualizing functionallity is usually more important than realism.
For virtual design reviews of actual machines, factories, or buildings, the underlying data is not created for the visualization itself but is based on potentially highly complex 3d CAD-data.
Converting such CAD data into a suitable virtual scene typically requires expert knowledge, manual work and a substantial amount of time.

We present a new approach for rendering very large and complex scenes supporting a wide range of input scenes, including CAD data, fast enough for displaying on HMDs.
Our approach is fast, robust, easy to implement, requires only minimal user interaction for preparing an input scene, and offers good visual quality, which automatically adapts to the required framerate and available GPU computation power.
Because our method does not have any requirements for the type of scene (e.g. landscape, architecture, machines), it can render any scene equally well.
No time-consuming manual work for converting CAD data is required.
This is achieved by combining and extending ideas from image-based and point-based rendering, visibility algorithms and approximation algorithms.

\subsection{Outline} 
The basic idea of our technique is to approximate complex parts of the scene having a small projected size by a set of points with much lower complexity.
In contrast to other point based approximation algorithms, the points are not placed on all surfaces, but only on surfaces that are visible if the approximated part is seen from the outside (external visibility).
To minimize the number of points needed to cover a surface without visible holes, the placement algorithm aims at maximizing the minimal distances between neighboring points (and thereby aims at creating a blue noise distribution).
Unlike other techniques distributing points evenly on a three dimensional surface, our algorithm creates a particular ordering of the distributed points:
Each prefix of the complete set of points of  an approximated part maximizes the closest distances between neighboring points.
Choosing a larger prefix results in a smaller distance between points and in a denser coverage of the surface.
This allows to dynamically choose as few points for a part of the scene as are necessary to cover each of its projected pixels with high probability.
The sorted surface points are created in a preprocessing step and are stored in ordinary vertex buffers.
During runtime a traversal algorithm determines which parts of the scene are rendered using the original geometry and which ones are approximated using a subset of points.
The number of rendered points is determined by the available projected size of the rendered part and the overall rendering budget for that frame.
A huge benefit of this arrangement is, that rendering one array of points requires only a single draw call to the GPU with a variable length parameter.
To our knowledge, other current progressive simplification methods still need to perform complex operations for the dynamic simplification or refinement on the CPU or GPU, or require a certain structure of the simplified object.

Our technique consists of two steps: In a preprocessing step, the surfels are generated based on the scene's original geometry.
We describe the generation of the inital surfel set, the sampling process and the surfel generation for hierarchical scenes in \autoref{sec:preprocessing}.
The second step is to render the precalculated surfel approximations during real-time rendering.
We present the rendering in \autoref{sec:rendering}.
In \autoref{sec:eval} we present experimental results evaluating the overhead of the preprocessing and rendering time and visual quality of the rendering for a highly complex virtual scene.
In \autoref{sec:conclusion} we discuss limitations and possible extensions to the technique.

\section{Related work}
\label{sec:related}

A lot of research has gone into the area of rendering highly complex 3d scenes in real time and vast amount of techniques has been developed over the years.
The usual approach is to reduce the amount of data that has to be processed by the graphics hardware by culling invisible parts of a 3d scene and reducing the complexity of some objects by replacing it with a simplification (level-of-detail) where feasible.

For level-of-detail (LOD) based algorithms the 3d scene is usally partitioned into a hierarchical data structure where each level provides an approximation of the underlying subtree (HLOD \cite{Erikson2001}).
The approximations can consist of a discrete set of geometric models with varying complexity \cite{Luebke2003}, image-based simplifications \cite{Aliaga1999,Oliveira2000,Sues2010}, point-based simplifications \cite{Gross2009,Kobbelt2004,Alexa2004}, or progressive level-of-detail approximations \cite{Hoppe1996,Yoon2004,Derzapf2012}.

Progressive LODs have the advantage, that the degree of abstraction can be chosen dynamically at run time dependent on the observers position and viewport, and therefore there is a continuous transition between different detail levels without \emph{popping} artifacts that occur when switching between discrete models.
Progressive Meshes were introduced by Hoppe \cite{Hoppe1996}.
A Mesh is progressively refined or simplified by performing a sequence of split or collapse operations on the vertices and edges on a mesh.
This idea was later combined with the idea of HLOD to allow for a hierarchiy of progressive simplifications \cite{Yoon2004}.
A problem of progressive meshes is, that they require a certain structure of the mesh (i.e. 2-manifold geometry), and do not translate well to the GPU, because of the dependencies between operations and vertices.

Although there are a few attempts for progressive meshes on the GPU \cite{Hu2009,Derzapf2012}, using progressive point-based approximations are often better suited, since they usually don't require neighborhood dependencies between points.
Dachsbacher et al. \cite{Dachsbacher2003} proposed a progressive point-based LOD technique that allows adaptive rendering of point clouds completely on the GPU.
They transfer points effectively to the GPU by transforming it into a sequential point tree which can be traversed directly on the GPU by sequential vertex processing.
A similar approach to progressive point rendering as progressive meshes by Hoppe \cite{Hoppe1996} was proposed by Wu et al. \cite{Wu2005}.
From an initial point set they arrange all possible splat merge operations into a priority queue according to some error metric.
The operations can then be iteratively performed to achieve the desired detail level.

To our knowledge, all available progressive LOD techniques require some sort of complex operations on the CPU or GPU to refine or simplify the geometry or point cloud.
We propose an approach that does not require any refinement or simplification operations.
We order the points in a vertex buffer such that each prefix presents a \emph{good} approximation and the detail level can be chosen by simply specifying the number of points to be rendered.
This can be achieved by ordering the points by an approximate \emph{greedy permutation} or \emph{farthest-first traversal} \cite{Eppstein2015}, a well known technique from image sampling \cite{Eldar1997}.
Using a farthest point strategy, Eldar et al. \cite{Eldar1997} showed that such a sampling scheme possesses good \emph{blue noise} characteristics.
Having a \emph{blue noise} quality for a set of point samples (either 2d or 3d) is a desirable property.
It guarantees high spatial uniformity and low regularity of the distribution which avoids aliasing artifacts when displaying the points \cite{Yan2015}.
Moenning and Dodgson \cite{Moenning2003a} used a farthest point strategy (FastFPS) for the simplification of point clouds and they also hinted at the usefulness of this strategy for progressive point rendering.
However, their algorithm requires a dense point cloud or an implicit surface as input.

For the rendering of point clouds, the most common practice is the usage of surfels or splats as introduced by Pfister et al. \cite{Pfister2000} and Rusinkiewicz et al. \cite{Rusinkiewicz2000}.
A surfel is a n-tuple which encodes a 3d position together with an orientation and shading attributes that locally approximate an object's surface.
The point cloud can then be rendered using hardware accelerated point sprites \cite{Coconu2002,Botsch2003} or screen space splatting \cite{Zwicker2001,Guennebaud2004a,Preiner2012}.
Multiple techniques have been developed over the years to improve the visual quality of the rendered surfels, and this is still an active research area.

One major difference of available point-based rendering techniques to our method is how we aquire the initial point set which then gets refined.
We only sample the externally visible surfaces of an object by rendering it from multiple directions and using the rendered pixels as basis for the surfel computations.
A similiar approach for sampling visibility was proposed by Eikel et al. \cite{Eikel2013}.
They use the rasterization hardware to compute the direction-dependent visibility of objects in a bounding sphere hierarchy.
This allows for efficient rendering of nested object hierarchies without the need of time consuming occlusion culling or high memory consumption for precomputed visibility.

\section{Generation of Surfel Approximations}
\label{sec:preprocessing}

In the following, we describe the generation of surfel approximations (LODs) for complex scenes.
A surfel approximation is stored in a single contiguous vertex buffer object where each vertex entry represents a surfel, where a surfel consists of a 3d position, a normal vector and material properties (e.g., color).
The order of the surfels in a vertex buffer gives an approximation of the underlying object that can be progressively refined by simply adjusting the number of rendered surfels.
This allows for an efficient, cache-friendly rendering of a surfel approximation independent of the desired detail level, by simply rendering only a prefix of the a single vertex buffer object.

We assume the scenes to be represented by a hierarchically organized scene graph, preferably representing the spatial structure of the scene.
Scenes originating from CAD data often already provide a suitable structure (object hierarchies, assembly groups).
If no such structure is available, commonly used spatial data structures can be applied, e.g. a loose octree \cite{Ulrich2000}.
We assume that the scene's geometry is stored in the leaf nodes of the scene graph and that the geometry is represented by polygonal data, although any renderable opaque surface representation can be used.

We begin by describing the generation of a surfel approximation for a single object, i.e., a single node in the scene graph.
First, we generate an initial set of surfel canditates, wich is described in \autoref{sec:pre:initial}.
Then, we progressively sample the initial set of candidates to achieve the desired surfel approximation of the object (\autoref{sec:pre:sampling}).
Finally, we describe the hierarchical generation of surfel approximations for an entire scene graph (\autoref{sec:pre:hierarchy}).

\subsection{Creating the initial set of surfels}\label{sec:pre:initial}

\begin{figure}
	\begin{center}
			\includegraphics[width=1.0\columnwidth]{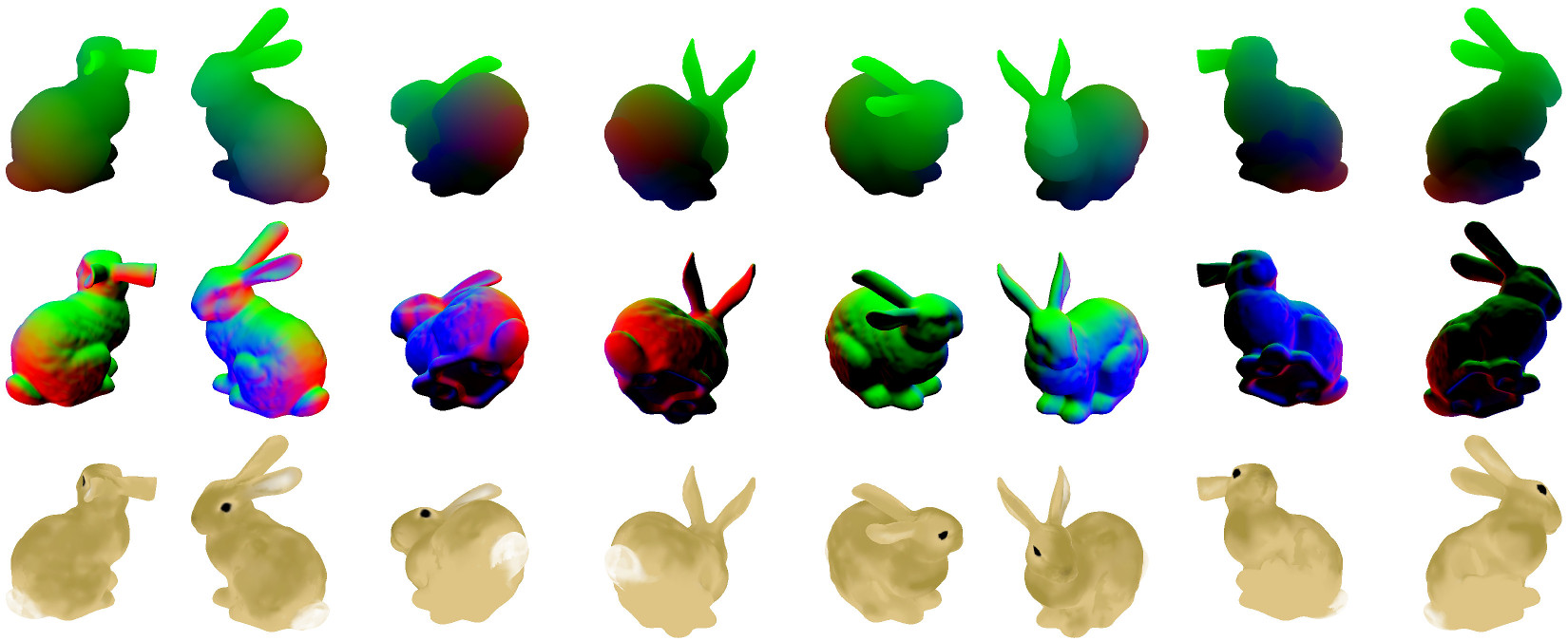}
	\end{center}
	\caption{Render buffers for 8 directions with position, normal, color}
	\label{fig:pre:buffers}
\end{figure}

The first step for creating the surfel approximation for a single node in a scene graph is to determine an \emph{initial set of possible surfels} from which the resulting surfels for the node's approximation will be drawn.
We generate these initial samples using the rasterization hardware by rendering the node's subtree from multiple directions into a set of G-buffers.
This allows us to capture the visible surface of a subtree as seen from outside of it's bounding volume.
In practice, it has shown that rendering from the eight corners of the node's bounding box, using orthographic projection directed to the center of the node, gives a sufficiently good approximation of the visible surface for most applications.
We use multi-target rendering to render the node into multiple output buffers in a single rendering pass, each output buffer encoding a different surface property.
One buffer contains the pixels' 3d position relative to the node's local coordinate system, another buffer encodes the surface normal in the same coordinate system.
At least one buffer is used to encode the pixels material properties.
In the simplest version, the ambient, diffuse and specular color values are combined into a single color value.
To allow for more complex lighting of the approximation during rendering, further properties can be captured in additional buffers, like PBR material parameters.
As this step uses the standard rendering pipeline, even complex shaders for surface modification and generation can be used if they provide the desired output and only produce opaque surfaces.
To assure a sufficiently dense coverage of the surface with surfels, the resolution for rendering the buffers for one direction should at least match the intended resolution used at runtime.
A much higher resolution unnecessarily increases the preprocessing time and requires a larger sample size during the following adaptive sampling phase.
An example for the created buffers for a single object is shown in \autoref{fig:pre:buffers}.
After the buffers have been filled, a surfel entry (a record storing a 3d position, normal, color etc.) is added to the initial set of surfels for each covered pixel.

\subsection{Progressive Sampling}\label{sec:pre:sampling}
After creating the initial surfel set, we select and sort a subset of the surfels.
The goal is to achieve a sequence of surfels in which for each prefix, the minimal closest distance between neighboring surfels is maximized (greedy permutation), i.e. the first surfels are evenly spread over different parts of the object while further surfels start covering the complete surface until small details are refined at the end of the sequence.
In order to approximate such a sequence, we apply a random sampling technique:
We start by selecting and removing a random starting surfel from the input set (the initial set of surfels) and add it to the output sequence.
Now, we select a uniformly chosen random subset of the remaining input set of fixed size (in practice, ${\sim}200$ samples yield a reasonably good quality result).
We determine the candidate with the largest distance to all surfels chosen before (e.g. using an octree data structure over the chosen surfels), append it to the result sequence, and remove it from the input set.
Samples that are at close vicinity to the chosen samples can also be removed from the input set.
The other samples remain in the input set for the next round.
The sampling step is repeated until a desired number of chosen surfels is reached or the input set becomes empty.
The number of created surfels influences the point size that can be chosen during rendering of the node and therefore the quality of the approximation.

In order to speed up the sampling process, we apply a heuristic:
After each $k$ iterations, the number of candidates chosen from the random sample set is increased by one (e.g., $k=500$).
This increases the prepocessing speed while reducing the quality of the distribution only slightly as shown in \autoref{sec:pre:distr}. 

\subsection{Quality of Sampling Distribution}\label{sec:pre:distr}

\begin{figure}
  \centering
  \smaller
  \newcommand{\rot}[1]{\rotatebox[origin=c]{90}{#1}}
  \newcommand{\img}[1]{\raisebox{-.5\height}{\includegraphics[width=0.255\columnwidth]{#1}}}
  \begin{tabular}{ccccc}
    \rot{Greedy Permutation} &  \img{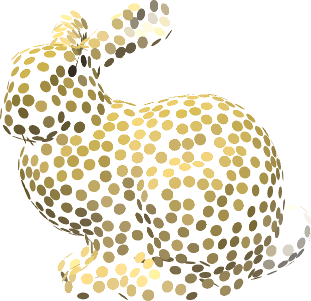} &  \img{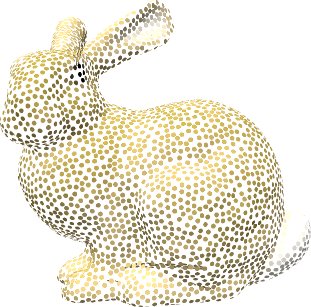} &  \img{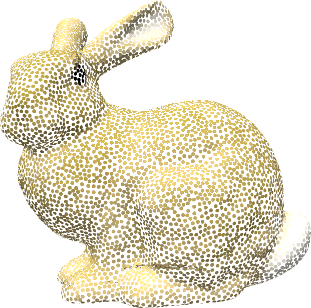} & \\
          \rot{Blue Surfels} & \img{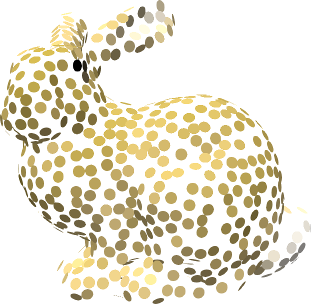} & \img{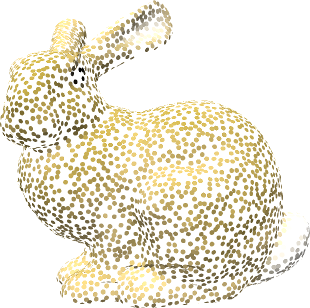} & \img{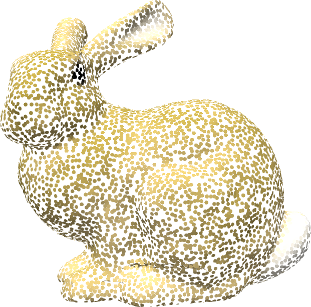} & \\
                \rot{Random} & \img{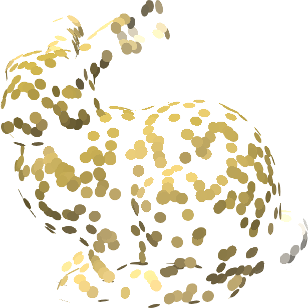} & \img{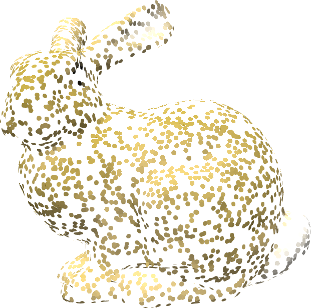} & \img{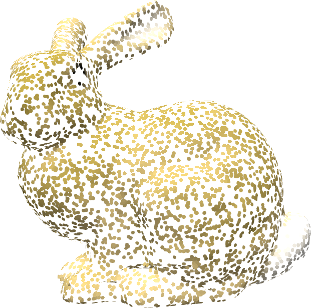} & \\
    \rot{Min. rel. point distances} & \multicolumn{4}{c}{\raisebox{-.5\height}{\includegraphics[width=0.89\columnwidth]{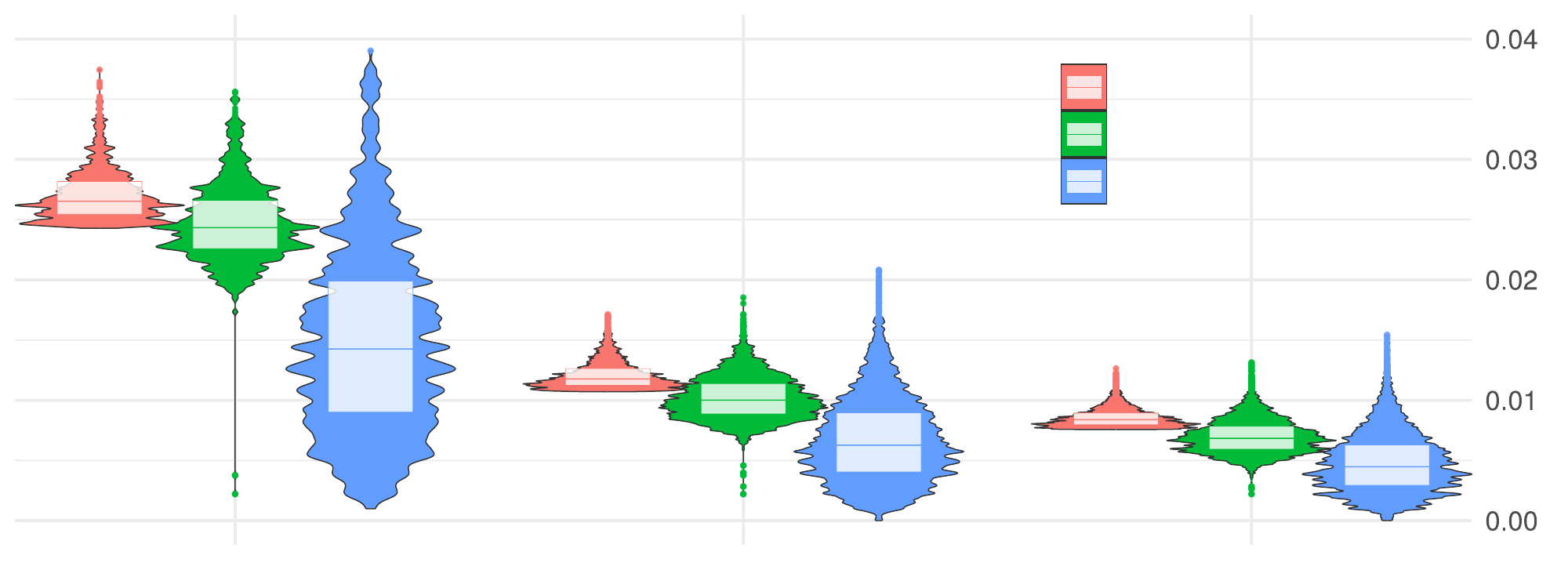}}} \\
    & 1k surfels & 5k surfels & 10k surfels
  \end{tabular}
	\caption{%
  {\textit Top 3 rows}: Different surfel prefixes (1k, 5k, 10k) for the stanford bunny model for different sample strategies. 
  Point sizes are reduced to show distributions.
  {\textit Below}: Violin plots for the minimum relative point distances for each prefix and strategy. 
  }
	\label{fig:pre:surfelPrefix}
\end{figure}

\begin{figure*}
  \centering
  \subfloat[100 surfels]{\includegraphics[width=0.2\textwidth]{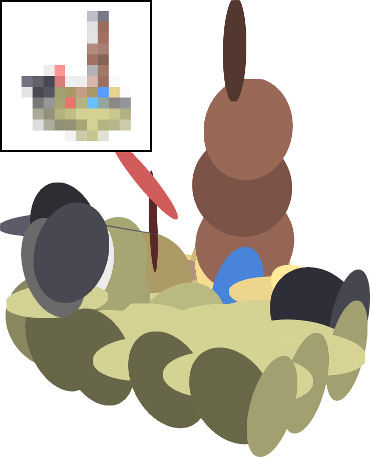}\label{fig:render:prefixes:100}}
  \subfloat[1000 surfels]{\includegraphics[width=0.2\textwidth]{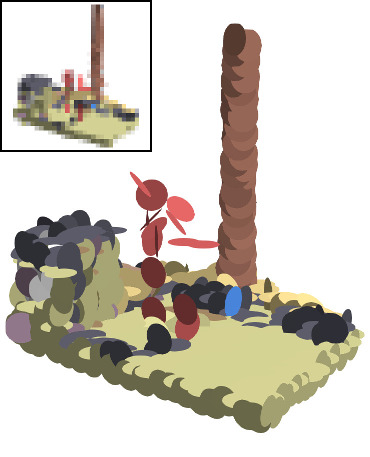}\label{fig:render:prefixes:1k}}
  \subfloat[10k surfels]{\includegraphics[width=0.2\textwidth]{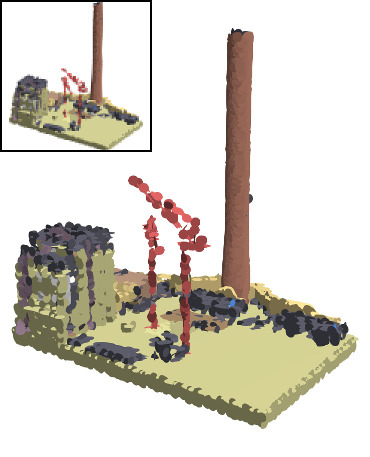}\label{fig:render:prefixes:10k}}
  \subfloat[100k surfels]{\includegraphics[width=0.2\textwidth]{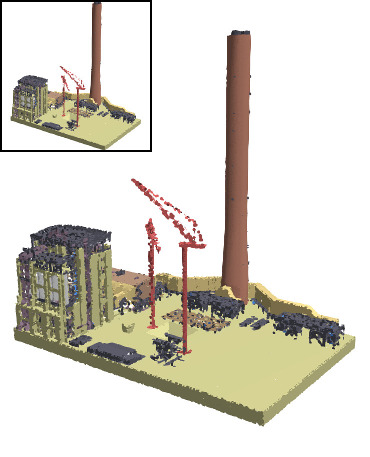}\label{fig:render:prefixes:100k}}
  \subfloat[original]{\includegraphics[width=0.2\textwidth]{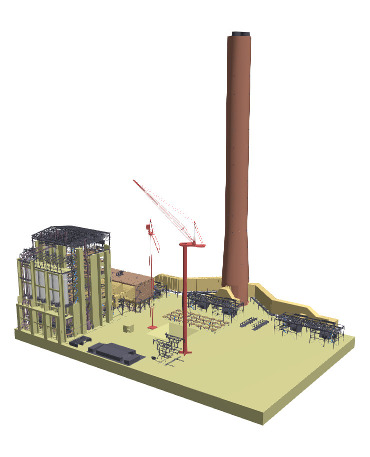}\label{fig:render:prefixes:orig}}
	\caption{%
  Illustration of different prefix lengths (100, 1k, 10k, 100k) of a single surfel approximation of the power plant model. 
  The boxes in the upper left corners of (a)-(d) show the surfel approximations rendered at their intended resolutions, while (e) shows the power plant model rendered without any approximation. 
  }
	\label{fig:render:prefixes}
\end{figure*}

Since the goal is to cover the entire surface of an object with as few surfels as possible to get the best possible image quality when rendering a surfel approximation, a uniform distribution of the points is very important.
A greedy permutation (a sequence of points where every point is as far away as possible from all previous points) has the property, that the points of every prefix are uniformly distributed with blue noise characteristics which reduces aliasing artifacts when displaying the points and therefore gives a good approximation of the underlying surface.
We use a simple randomized sampling algorithm, which runs indepentently from the size of the input data, to quickly get an approximate greedy permutation.
Although we might loose some of the desired properties, our experiments show, that our method still gives a reasonably good approximation of a greedy permutation while perfoming much faster than an exact greedy permutation (computing an exact greedy permutation for 10k points of the bunny model shown in \autoref{fig:pre:surfelPrefix} took us ${\sim}15s$ while our method only took ${\sim}180ms$).

\autoref{fig:pre:surfelPrefix} shows surfel prefixes of different sizes ($1k, 5k, 10k$) for the stanford bunny model\footnote{http://graphics.stanford.edu/data/3Dscanrep/} in comparison to surfels chosen uniformly at random and using an exact greedy permutation (based on the initial surfel set).
Although our method lacks in the good uniformity of the greedy permutation, it still proves a significant improvement over the random solution.
This can be especially seen at smaller prefix lengths of $1k-5k$ surfels.
While the random solution forms many clusters of surfels as well as holes, our solution is only slightly different from the exact solution.
The difference between our solution and the exact solutions becomes more visible at $10k$ surfels.
However, the points are still roughly uniformly distributed on the surface.

The bottom row of \autoref{fig:pre:surfelPrefix} shows a combined violin and box plot of the minimum relative point distances between surfels at the given prefix sizes and for each of the three sampling methods.
Since the greedy permutation maximizes the minimum distance between points for each prefix, one can see a clear minimum cap at some distance while our method has more outliers that fall below this value.
However, the median distance of our method is still close to the minimum distance of the greedy permutation approach while significantly better than the random approach.
This is an indicator that our method yields a good surface coverage for a fixed prefix length and therefore allows for fast rendering with a good image quality in comparison to the other two solutions.
The image quality is further examined in \autoref{sec:eval}.

\subsection{Hierarchical generation}\label{sec:pre:hierarchy}

For the rendering of complex scenes, we hierarchically generate surfel approximations for multiple subtrees of the existing scene graph structure of a scene.
This can be done by traversing the scene graph in a top-down or bottom-up order and generate surfel approximations for each node that exceeds a certain complexity (e.g., generate a surfel approximation of $10k-100k$ surfels for each subtree that consists of more than $10k$ triangles).
When generating the surfel approximations bottom-up, one can use the already computed surfels of the child nodes instead of the original geometry to speed up the rendering step for the initial surfel sampling (see \autoref{sec:pre:initial}).
Otherwise, any existing approximation or culling technique can be used to generate the images for the initial sampling process.
This also allows for easy out-of-core generation of the surfel sets.

For animated or moving objects, one should generate the surfel approximations seperately from the static scene parts, since already computed approximations cannot easily be modified without breaking the desired distribution qualities.
Unfortunately, complex deforming animations cannot easily be handled by our method, but it shouldn't be too difficult to incorporate bone weights for skeletal animation in the vertices of the surfel buffer.

\section{Rendering Progressive Blue Surfels}\label{sec:rendering}

In this section, we describe the rendering of Progressive Blue Surfels during an interactive walkthrough of a complex scene.
The goal is to replace entire subtrees of a scene graph with their corresponding surfel approximation (LOD) as long as the visible surface of the original geometry can be covered by the oriented discs defined by the surfels and as long as the image quality (and run time) suffices for the intended application.
Given a fixed surfel size, we can easily compute the required prefix of the surfel approximation dependent on the distance of the approximated object to the observer to cover all pixels of the object in screen space (see~\autoref{sec:rendering:prefix}).
An example of different prefix length  (100, 1k, 10k, 100k) of a blue surfel approximation of the UNC power plant model \cite{WalkthruGroup2001} can be seen in \autoref{fig:render:prefixes} compared to the model without any approximation (\autoref{fig:render:prefixes:orig}).
It also shows a zoomed in view (upper left boxes of Figure~\ref{fig:render:prefixes:100}-\ref{fig:render:prefixes:100k}) for each of the surfel prefixes as they would actually be rendered with the corresponding prefix length.

We render the surfels as oriented discs by using \emph{OpenGL} point primitives together with a fragment shader as described in \autoref{sec:rendering:ellipses}.
When a surfel approximation for a scene node cannot sufficiently cover the visible geometry of the subtree anymore (i.e., when getting too close to the object), we blend between the node's approximation and its children's approximations or original geometry by gradually decreasing the number of rendered surfels for the node while increasing the number of rendered surfels of the child nodes or rendering the original geometry.
Finally, in \autoref{sec:rendering:adaptive}, we describe a simple extension to our rendering algorithm that adaptively tries to keep a desired frame-rate while maximizing the possible image quality, and in \autoref{sec:rendering:foveated}, we describe how our method can be used for simple fixed foveated rendering for head-mounted displays. 

\subsection{Rendering a surfel prefix}\label{sec:rendering:prefix}
A surfel approximation for a single object is stored in a contiguous vertex buffer on the GPU.
We can choose the quality of the approximation by simply adjusting the number of rendered point primitives from this buffer.
Ideally, we choose the rendered prefix of the buffer such that the entire surface of the object is covered by surfels without holes, i.e., we choose the prefix length in such a way, that every pixel of the rendered object is covered by at least one surfel.
That means, for a given surfel radius $r$, we want to find a minimal prefix length $p$ s.t. every other surfel in the entire surfel set (which approximates the surface) is covered by a surfel of radius $r$ in the prefix.
To find this value $p$, we use the close relation of greedy permutations to $r$-nets.
An $r$-net of a point set $X$ is a subset $S\subset X$ s.t. no point in $S$ is within a distance of $r$ of each other and every point in $X$ is within a distance of $r$ to a point in $S$.
Now, each prefix of a greedy permutation is an $r$-net for $r$ equal to the minimum distance between points in this prefix \cite{Eppstein2015}.
Using this, and the fact, that for Euclidean metrics in $\mathbb R^d$, every ball of radius $r$ can be covered by $\Theta(2^d)$ balls of radius $r/2$ \cite{Gupta2003}, we can easily estimate the prefix length~$p$ for a given radius~$r$:
\[p(r)=p_m\cdot\left(\frac{r_m}{r}\right)^2\]
Here, $r_m$ is the precomputed median minimum distance for a fixed prefix length $p_m$ (e.g., $p_m=1000$), which can easily be computed during the preprocessing phase (see \autoref{sec:pre:sampling}).
We use the median of the minimum distances between points to compensate for our used heuristic.
We furthermore use the simplified assumption, that the generated surfels lie on a 2-manifold surface (which does not have to be the case), i.e. every surfel disk covers ${\sim}2^2=4$ surfels with half the radius (in $\mathbb R^2$ this value should be ${\sim}7$, but $4$ gives a good enough estimation).

To get a covering of all pixels in screen-space, we choose $r$ proportional to the projected distance between two pixels relative to the approximated object's local coordinate system.
For this, we take two neighboring pixels in screen-space, project them onto the view plane going through a point at the object (e.g., the closest point to the bounding box of the object from the camera position) and compute the distance $d_p$ between these two points (in the object's local coordinate system).
Now, we compute $r$ by $r=\frac{s}{2\cdot d_p}$, where $s$ is the desired surfel size in pixels (which corresponds with \verb|gl_PointSize| in \emph{OpenGL}).

\subsection{Drawing oriented discs}\label{sec:rendering:ellipses}
We render each surfel of a surfel prefix as a point primitive with a fixed size, i.e., a square in screen space with fixed pixel dimensions.
To draw the points as oriented discs (using the stored normals of a surfel), we use a fragment shader.
For each rendered fragment of a point primitive, we project the fragment in screen space back onto the plane defined by the surfels position and normal.
The fragment is discarded if the distance to the center of the surfel is larger than the size of the surfel in object space.
This results in opaque elliptic discs as can be seen in \autoref{fig:render:prefixes}.

In this step, it is also possible to use extended filtering methods for surfels as, e.g., EWA filtering as described by Botsch et al. \cite{Botsch2005}, to blend between the colors of neighboring surfels.
However, when rendering massive scenes using our method, such filtering methods can become too slow very quickly and
for our purposes, it was enough to only render elliptical splats without further filtering.
But, as shown in \autoref{sec:eval}, we still achieve a reasonable good image quality.

\begin{figure*}[t]
	\begin{center}
			\includegraphics[width=0.99\textwidth]{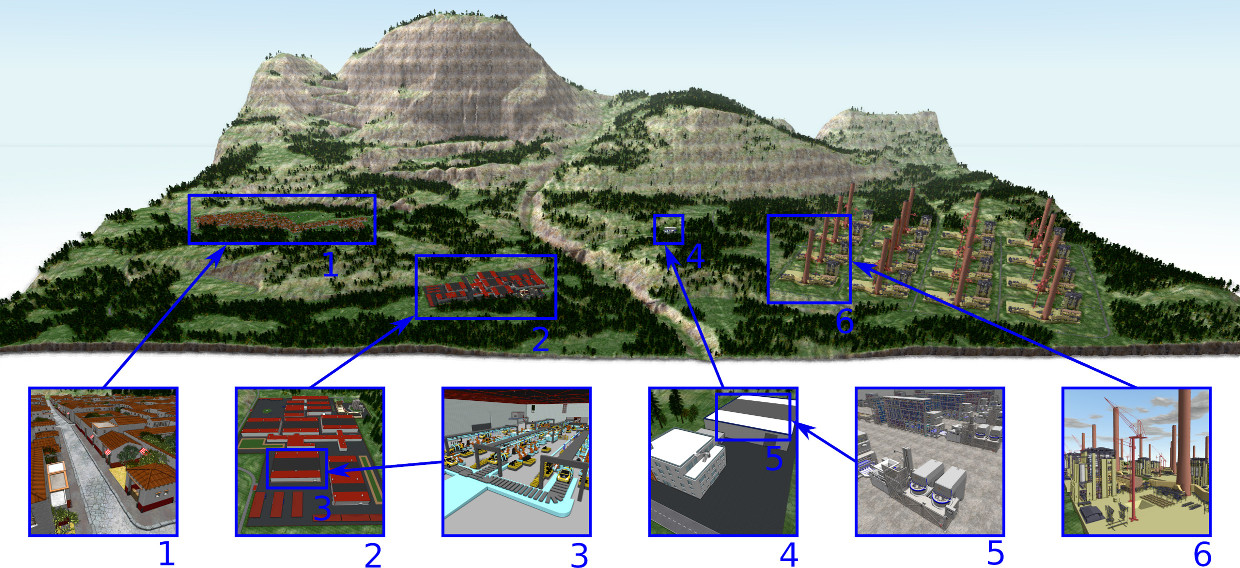}
	\end{center}
	\caption{Overview of the scene used for evaluation with highlighted scene parts. 1 - Pompeii; 2,3 - Car factory; 4,5 - Bakery; 6 - Power Plants}
	\label{fig:eval:overview}
\end{figure*}

\begin{table*}[t]
  \caption{
  Number of Objects, LODs (surfel approximations), Triangles, Surfels, and memory consumption of each scene part.
  The numbers show the total number including instanced geometry (objects that share the same memory), while the numbers inside parentheses show the number of unique geometry (without instancing).}%
  \label{fig:eval:scene}
	\centering
  \smaller
	\begin{tabular}{|r|r|r|r|r|r|r|r|} \hline
  	\textbf{Scene part} & \textbf{Objects (unique)} & \textbf{LODs (unique)} & \textbf{Triangles (unique)} & \textbf{Surfels (unique)} & \textbf{Triangle Memory} & \textbf{Surfel Memory} & \textbf{Total Memory} \\\hline
                Terrain &             $30518~(524)$ &          $30259~(265)$ &           $1.74 G~(1.73 M)$ &        $1.76 G~(24.74 M)$ &        $59.11~\text{MB}$ &     $494.79~\text{MB}$ &    $553.90~\text{MB}$ \\\hline
                Pompeii &           $33167~(26878)$ &          $5547~(2415)$ &         $62.93 M~(40.35 M)$ &       $40.28 M~(29.01 M)$ &         $1.79~\text{GB}$ &     $580.26~\text{MB}$ &      $2.37~\text{GB}$ \\\hline
            Car Factory &              $1708~(179)$ &             $893~(62)$ &          $75.85 M~(3.97 M)$ &        $17.36 M~(2.13 M)$ &       $130.71~\text{MB}$ &      $42.60~\text{MB}$ &    $173.32~\text{MB}$ \\\hline
                 Bakery &              $7237~(511)$ &           $6432~(363)$ &        $267.80 M~(27.70 M)$ &      $111.26 M~(10.23 M)$ &       $857.13~\text{MB}$ &     $204.76~\text{MB}$ &      $1.06~\text{GB}$ \\\hline
           Power Plants &              $4560~(304)$ &           $2145~(143)$ &        $186.00 M~(12.40 M)$ &        $99.67 M~(6.64 M)$ &       $364.15~\text{MB}$ &     $132.90~\text{MB}$ &    $497.05~\text{MB}$ \\\hline
                  Total &           $77495~(28397)$ &         $45419~(3248)$ &          $2.35 G~(86.23 M)$ &        $2.04 G~(72.76 M)$ &         $3.20~\text{GB}$ &       $1.45~\text{GB}$ &      $4.66~\text{GB}$ \\\hline
	\end{tabular}
\end{table*}

\subsection{Adaptive rendering}\label{sec:rendering:adaptive}
Due to the progressive nature of our method, it is easy to extend our algorithm for rendering complex scenes with an adaptive level-of-detail mechanism that tries to keep a desired frame-rate while maximizing the possible image quality.
The image quality of a surfel approximation depends mainly on the size of the rendered surfels (smaller is better) while the frame-rate depends on the number of rendered surfels and polygon count of the original geometry.
Now, for our method, the number of rendered surfels can be directly derived from the desired surfel size (or vice versa) to cover the visible surface of the original geometry (see \autoref{sec:rendering:prefix}).
Therefore, we can easily reduce the frame-time by increasing the size of the rendered surfels and therefore reducing the image quality.
This allows for a simple reactive algorithm that increases or decreases the surfel size based on the frame-time of the last rendered frame.
We do this by calculating the moving average of the surfel size $s_{old}$ of the last 3 frames and the surfel size weighted by the deviation factor of the last frame-time $t_{frame}$ to the target frame-time $t_{target}$:
\[s_{new}=\frac{s_{old}\cdot3+s_{old}\cdot t_{frame}/t_{target}}{4}\]
To avoid flickering, we only modify this value when the deviation factor reaches a certain threshold, e.g. when it falls outside of the interval $[0.9,1.1]$.
We also clamp the value by a minimum size of $1px$ and a small maximum size (e.g., $8px$) to avoid surfels that are too big.

\subsection{Foveated rendering}\label{sec:rendering:foveated}

To allow efficient rendering for head-mounted displays (HMDs), foveated rendering is a great method to significantly improve performance on current HMDs, even without eye-tracking capabilities.
The basic idea is, to decrease rendering complexity and quality in the periphery of the viewport while maintaining high fidelity in the focal area (fovea).
This is possible due to the distortion by the HMD lenses which skewes pixel at the border of the frame buffers for each eye.

With our method it is easily possible to implement a simple foveated rendering technique.
For this, we defined different fovea zones in screen space for each eye with different quality settings for the surfel sizes.
We then simply interpolate the size of the surfels between zones on a per-object basis to achieve a gradual increase in surfel size to reduce the complexity towards the periphery.
This allows for a smooth change in quality which is very important for rendering on HMDs since popping artifacts are especially noticable in the peripheral view.

\begin{figure}[hb]
	\centering
	\includegraphics[width=\columnwidth]{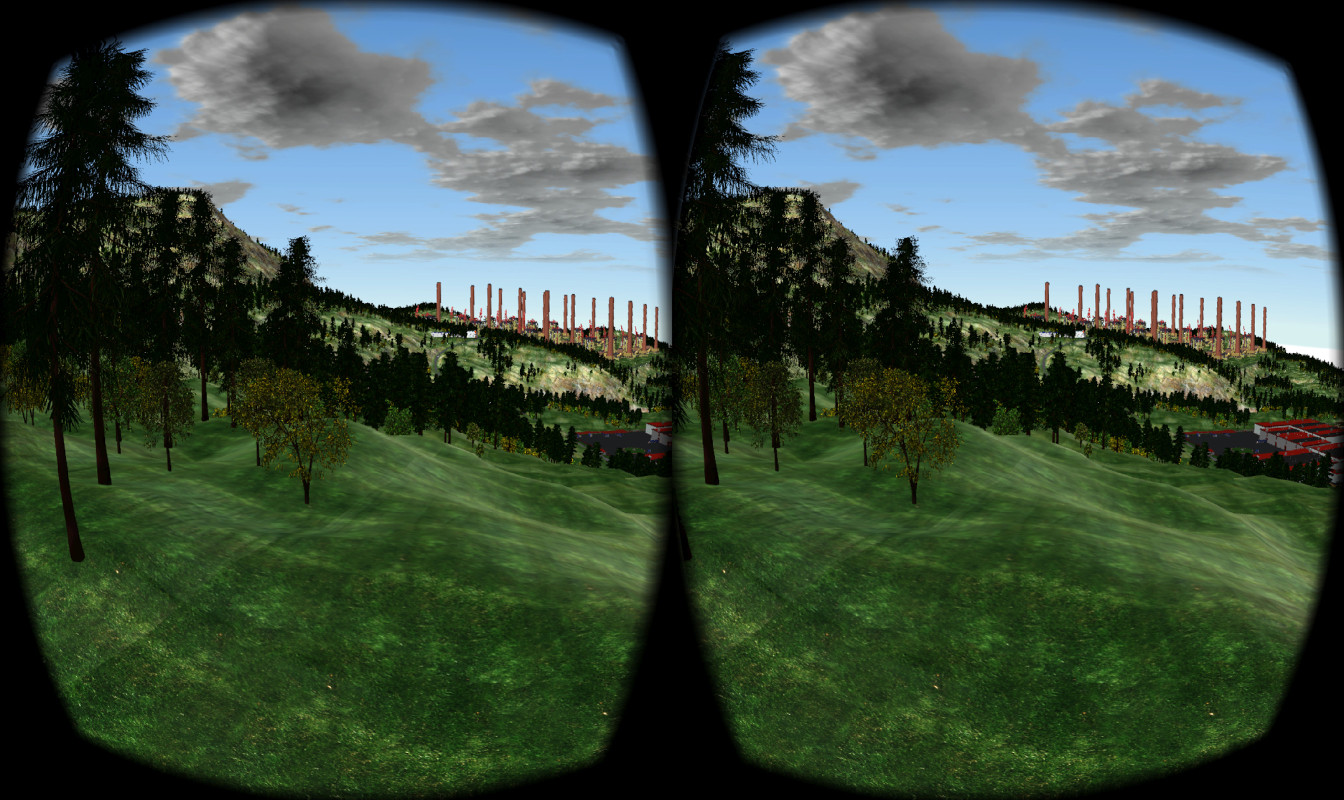}
	\caption{Example view of HMD stereo rendering on an Oculus Rift CV1 using our method.}
	\label{fig:rendering:hmd}
\end{figure}

\section{Results}\label{sec:eval}

In this section we describe the experimental results of our proposed rendering method.
In \autoref{sec:eval:benchmark} we describe the hardware configuration and the test scene used for evaluating the performance and visual quality of our method.
In \autoref{sec:eval:preprocess} we examine the preprocessing time of our method.
Subsequent, the running time and visual quality during rendering is discussed (\autoref{sec:eval:rendering}).

\subsection{Benchmark}\label{sec:eval:benchmark}

We implemented Progressive Blue Surfels in our experimental rendering framework.
All measurements of the subsequent evaluations were performed using a workstation PC (Intel Core i7-6700 with $4\times3.4$ GHz, 32 GB RAM, NVIDIA GeForce GTX 1060).
For experiments involving Head-Mounted Displays we used the Oculus Rift CV1 and stereo rendering using an oversized framebuffer of $2\by1344\by1600$.
An example image rendered using our method on the Oculus Rift can be seen in \autoref{fig:rendering:hmd}.

For our tests, we create a large heterogenous scene which consists of various parts (see~\autoref{fig:eval:overview}).
The basic scene is a $5km\times3km$ terrain chunk with roads and a high number of trees.
On this terrain we placed a set of smaller scenes that each highlight different strengths of our proposed rendering algorithm:
\begin{description}
  \item[Terrain] The terrain consists of $260$ tiles, each having a size of $255m\by255m$ and containing $4914$ Triangles (hexagonal grid). 
                 On the terrain we placed $30k$ trees (randomly selected from 6 unique trees with ${\sim}40k{-}70k$ triangles each).
                 The other scene parts are connected by roads consisting of $258$ simple road segments with ${\sim}100{-}5000$ triangles each.
  \item[Pompeii] Highly detailed model of pompeii generated using CityEngine\footnote{http://www.esri.com/software/cityengine/industries/procedural-pompeii} \cite{Mueller2006} (\autoref{fig:eval:overview} view 1).
                 It consists of a high amount of small objects with various materials.
  \item[Car Factory] A large car factory created during a student project consisting of multiple factory halls with moderately complex machinery and car parts (\autoref{fig:eval:overview} view 2\&3).
  \item[Bakery] A smaller factory hall with 5 high detailed triangluated CAD models of donut production lines provided by WP Kemper GmbH\footnote{https://www.wp-kemper.de} (\autoref{fig:eval:overview} view 4\&5). 
                Each production line contains ${\sim}50M$ triangles.
  \item[Power Plants] 16 copies of the UNC Power Plant model \cite{WalkthruGroup2001} with ${\sim}12M$ triangles each (\autoref{fig:eval:overview} view 6).
\end{description}

In total the scene consist of ${\sim}2.35G$ triangles in $77495$ individual objects ($28397$ unique triangle meshes).
A more detailed breakdown of the scene geometry and memory consumption of each part, as well as the generated surfel approximations, can be gathered from \autoref{fig:eval:scene}.

\subsection{Preprocessing}\label{sec:eval:preprocess}

\begin{figure}[htb]
	\centering
  \begin{tabular}{c}
	\includegraphics[width=0.9\columnwidth]{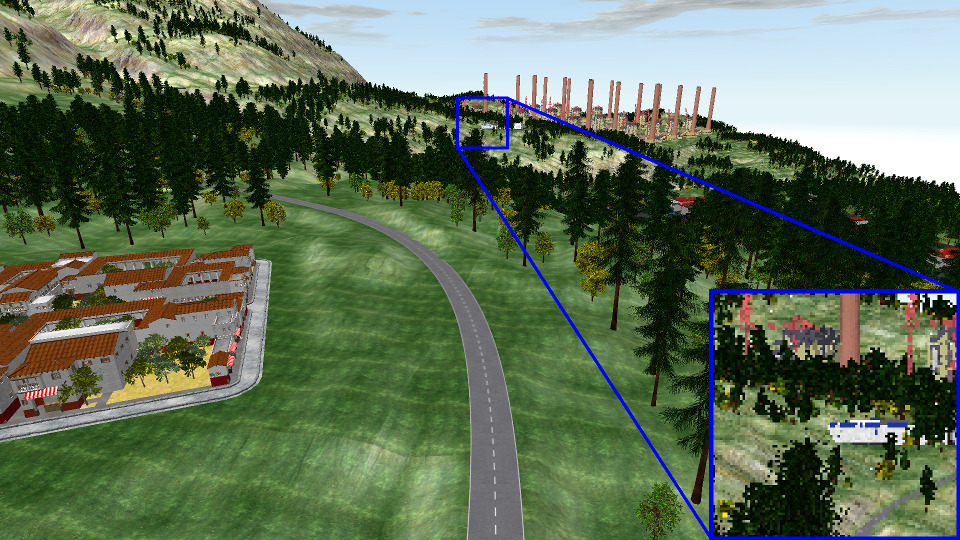}\\%
	\includegraphics[width=0.9\columnwidth]{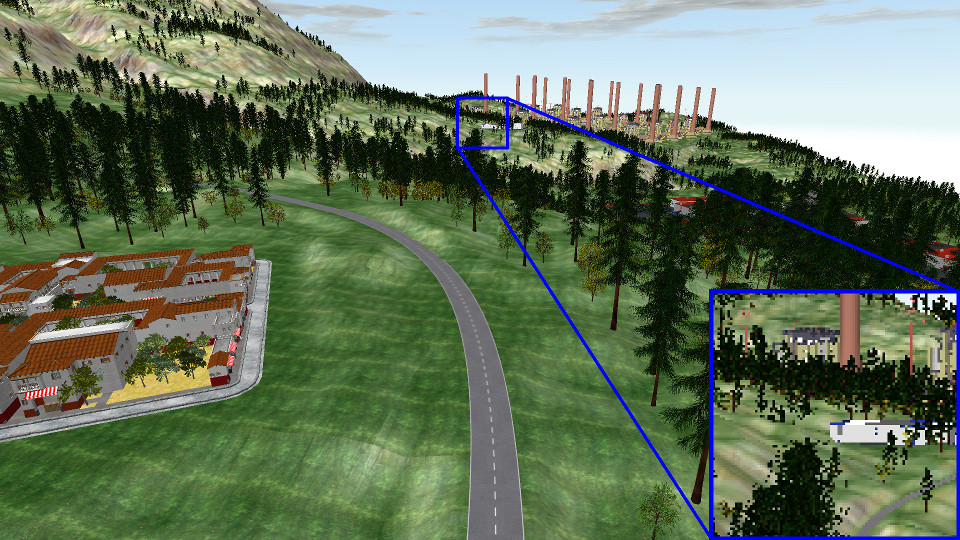}\\%
	\includegraphics[width=0.9\columnwidth]{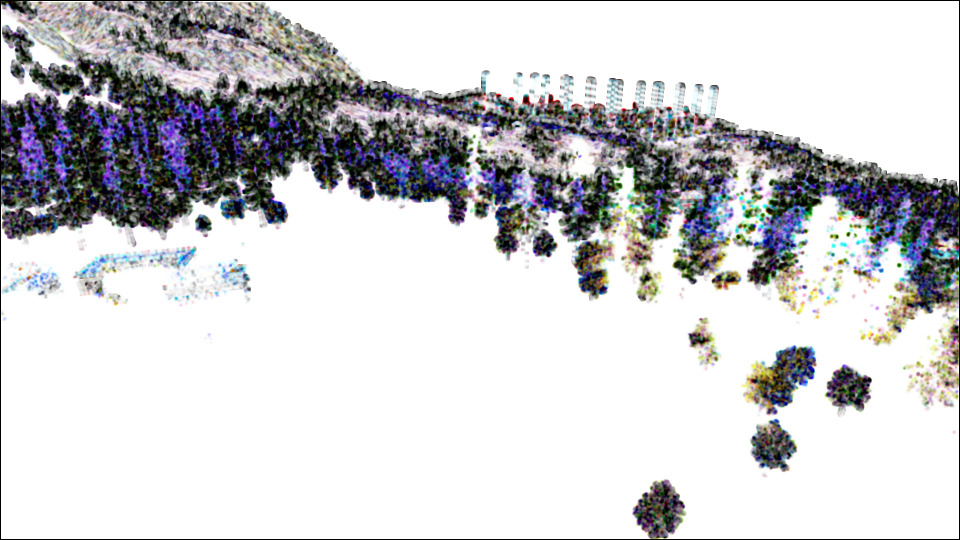}
  \end{tabular}
	\caption{%
    Comparision of a view rendered with our method (\textit{top}) and without LOD (\textit{middle}). 
    The \textit{bottom} image shows the image resulting from computation of the SSIM index showing the differences between the two images above.
    The resulting SSIM index is 0.81.
  }
	\label{fig:eval:quality}
\end{figure}

\begin{table}[t]
  \caption{Preprocessing times of the power plant model for varying surfel counts (10k, 50k, 100k).}%
  \label{fig:eval:single}
	\centering
  \smaller
	\begin{tabular}{|r|r|r|r|} \hline
                                  Surfel count &    10k &    50k &   100k \\\hline
                 Render to texture ($8\times$) &  12 ms &  12 ms &  12 ms \\\hline
    Creating initial surfel set (1.75M entries)&  82 ms &  82 ms &  82 ms \\\hline
                    Sampling (sample size 200) & 105 ms & 311 ms & 592 ms \\\hline
                                \textbf{Total} & 199 ms & 405 ms & 686 ms \\\hline
	\end{tabular}
\end{table}

In this section we examine the preprocessing time of our method.
\autoref{fig:eval:single} shows the preprocessing times for generating surfel approximations of various sizes for a single object (the UNC power plant \cite{WalkthruGroup2001}).
The initial samples were generated using a resolution of $1024\by1024$ using 8 directions.
For the progressive sampling, a sample size of 200 samples per round were chosen.
The only part that depends on the approximated scene geometry is the first rendering step from multiple directions from which the initial sample set is created.
However, this is usually only a small part for the generation and can be sped up by using preexisting culling or approximation techniques, or by using previously computed surfel approximations (when approximating larger subtrees of a scene graph).
The time taken for creating the initial surfel set is primarily due to transfering the data from the GPU to main memory and depends on the resolution used for creating the samples and (to some degree) the shape of the approximated object. 
The major portion of the computation is the sampling part, which (due to our randomized sampling technique) only depends on the intended target size of the surfel approximation.
This step can easily be multithreaded when generating LODs for an entire scene.

Our benchmark scene contains a total number 3248 unique surfel approximations of varying sizes ($1k{-}200k$ surfels).
We bounded the surfel count for an object by the minimum of $200k$ surfels and half the complexity of the approximated object (number of triangles).
Objects with a complexity of below 1000 triangles were only approximated in groups in a higher hierarchy level.
The total preprocessing time for the scene took only ${\sim}23.62$ minutes on a single thread. 

\subsection{Rendering Performance \& Image Quality}\label{sec:eval:rendering}

\begin{table*}[t]
  \caption{Rendering statistics of our method for different fixed camera positions and different resolutions including HMD stereo rendering. 
  The rightmost column shows the distribution of frame times in a combined box \& violin plot, measured at uniformly distributed points in a local area of the shown view. }%
  \label{fig:eval:performance}
  \newcommand{\tblimg}[2][2.25cm]{\raisebox{-.5\height}{\multirow{5}{*}{\includegraphics[height=#1]{#2}}}}
  \newcommand{\mr}[1]{\multirow{5}{*}{\rotatebox[origin=c]{90}{#1}}}
  \renewcommand{\arraystretch}{1.43}
	\centering
  \smaller  
  \begin{tabular}{|c|c|c|c|c|c|c|c|c|} \hline
        \# &               View &            &     No LOD &       1k &     2.5k &       4k &       HMD & Distribution of frame times \\\hline
    \mr{1 - Overview}     & \tblimg{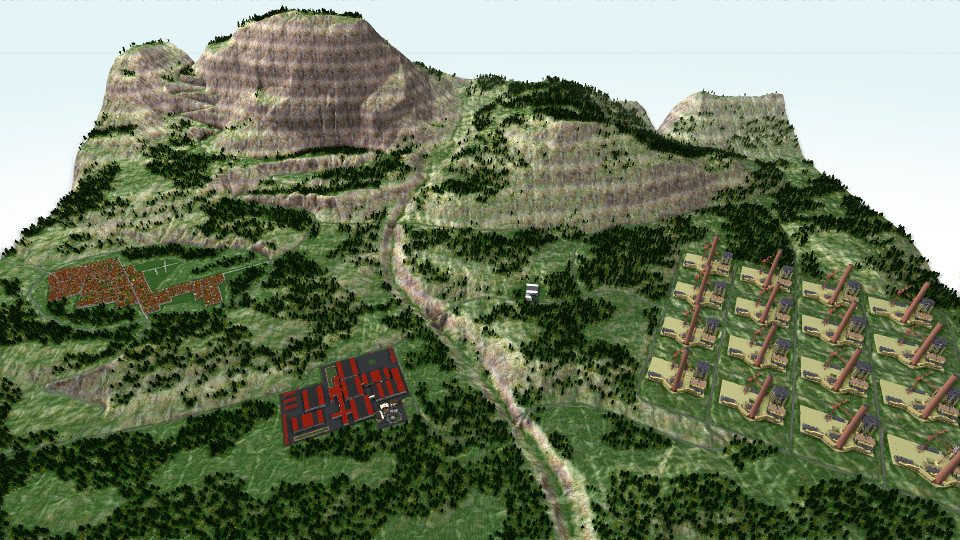} & Draw calls &    $73518$ &    $364$ &    $825$ &   $5513$ &     $682$ & \tblimg{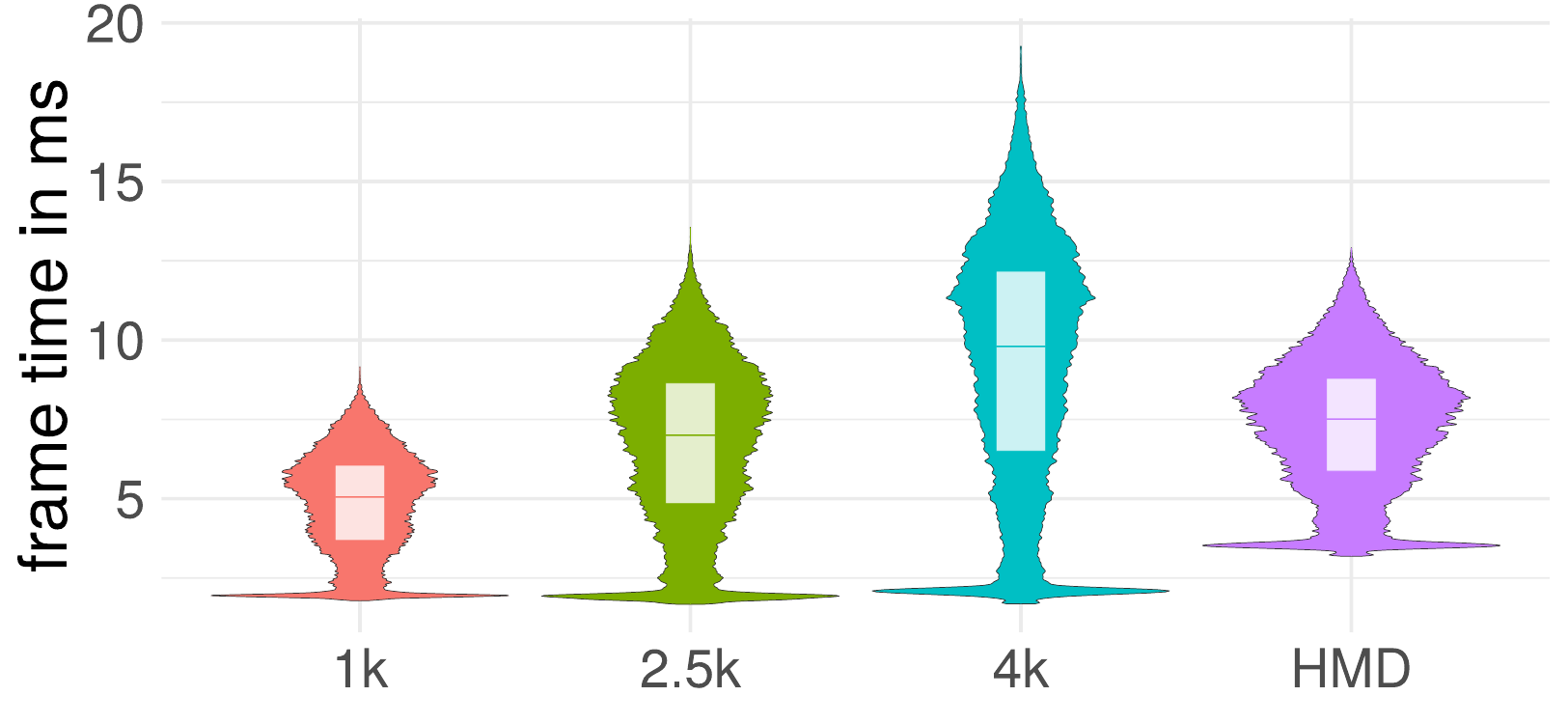} \\
                          &                    &  Triangles &    $2.12G$ & $92.22k$ &  $1.59M$ &  $1.86M$ & $179.45k$ & \\
                          &                    &    Surfels &        $0$ &  $4.32M$ &  $7.60M$ & $15.86M$ &   $4.97M$ & \\
                          &                    &        FPS &     $1.04$ & $271.95$ & $145.93$ &  $99.51$ &  $171.69$ & \\
                          &                    &    Quality &      $1.0$ &   $0.68$ &   $0.69$ &   $0.70$ &       $-$ & \\\hline
            
    \mr{2 - Pompeii}      & \tblimg{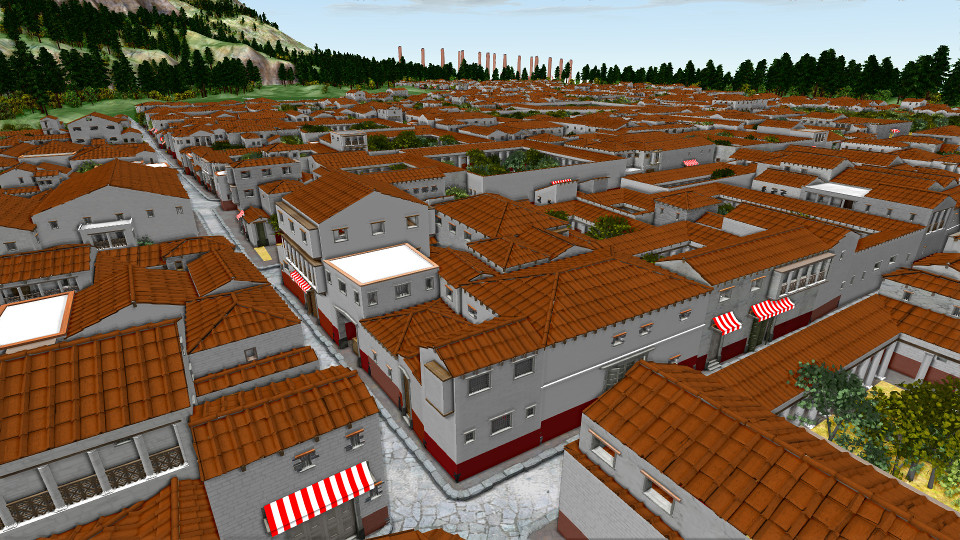} & Draw calls &    $56467$ &   $9770$ &  $12974$ &  $19943$ &   $17359$ & \tblimg{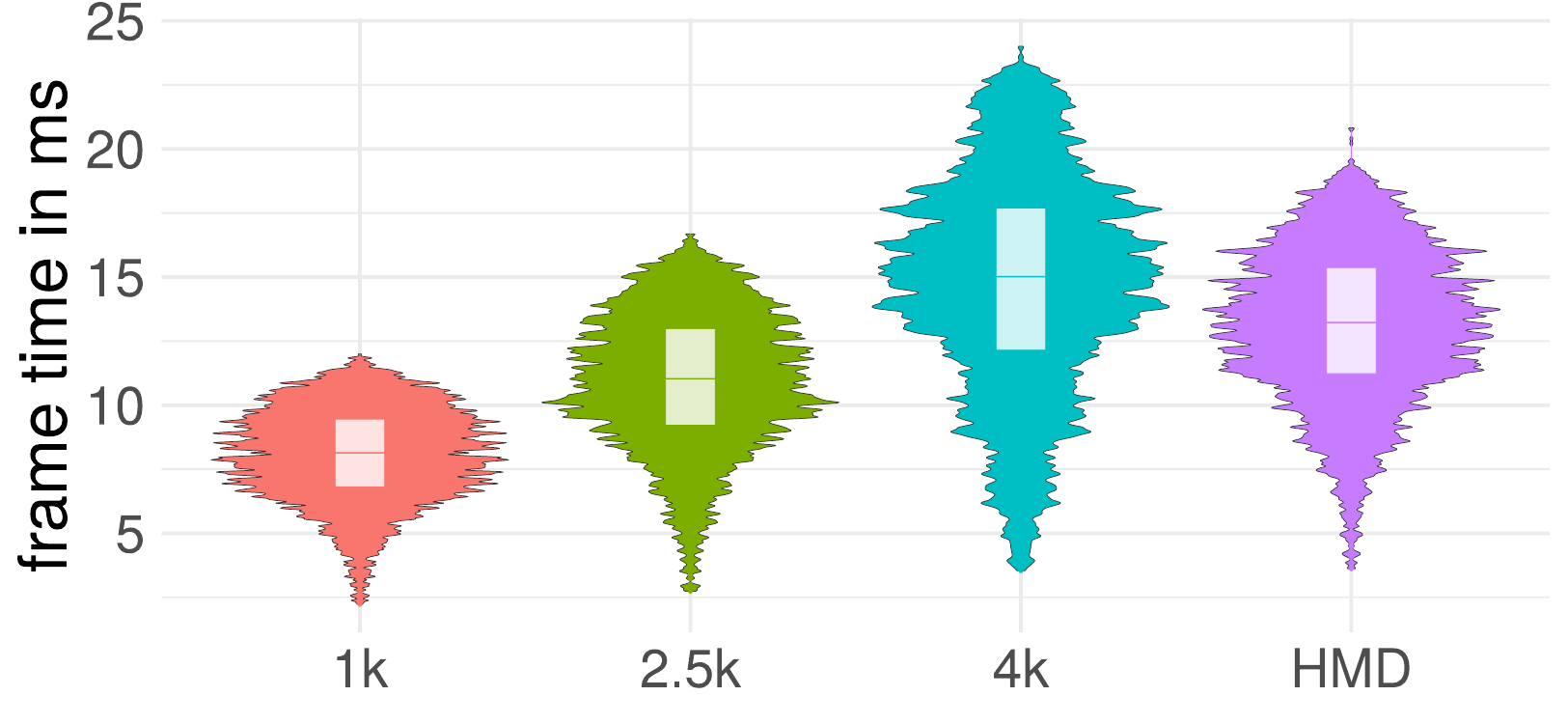} \\
                          &                    &  Triangles &    $1.73G$ & $11.61M$ & $15.36M$ & $25.53M$ &  $20.25M$ & \\
                          &                    &    Surfels &        $0$ & $12.00M$ & $17.10M$ & $27.58M$ &  $15.55M$ & \\
                          &                    &        FPS &     $1.31$ &  $81.02$ &  $54.49$ &  $43.57$ &   $43.71$ & \\
                          &                    &    Quality &      $1.0$ &   $0.92$ &   $0.93$ &   $0.94$ &       $-$ & \\\hline
            
    \mr{3 - Car Factory}  & \tblimg{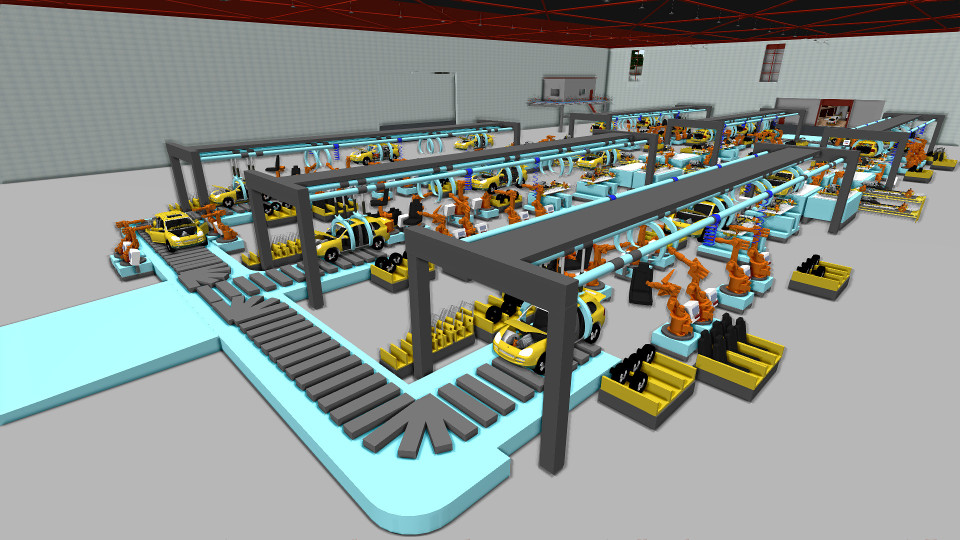} & Draw calls &    $12950$ &   $3641$ &   $4811$ &   $5748$ &    $6419$ & \tblimg{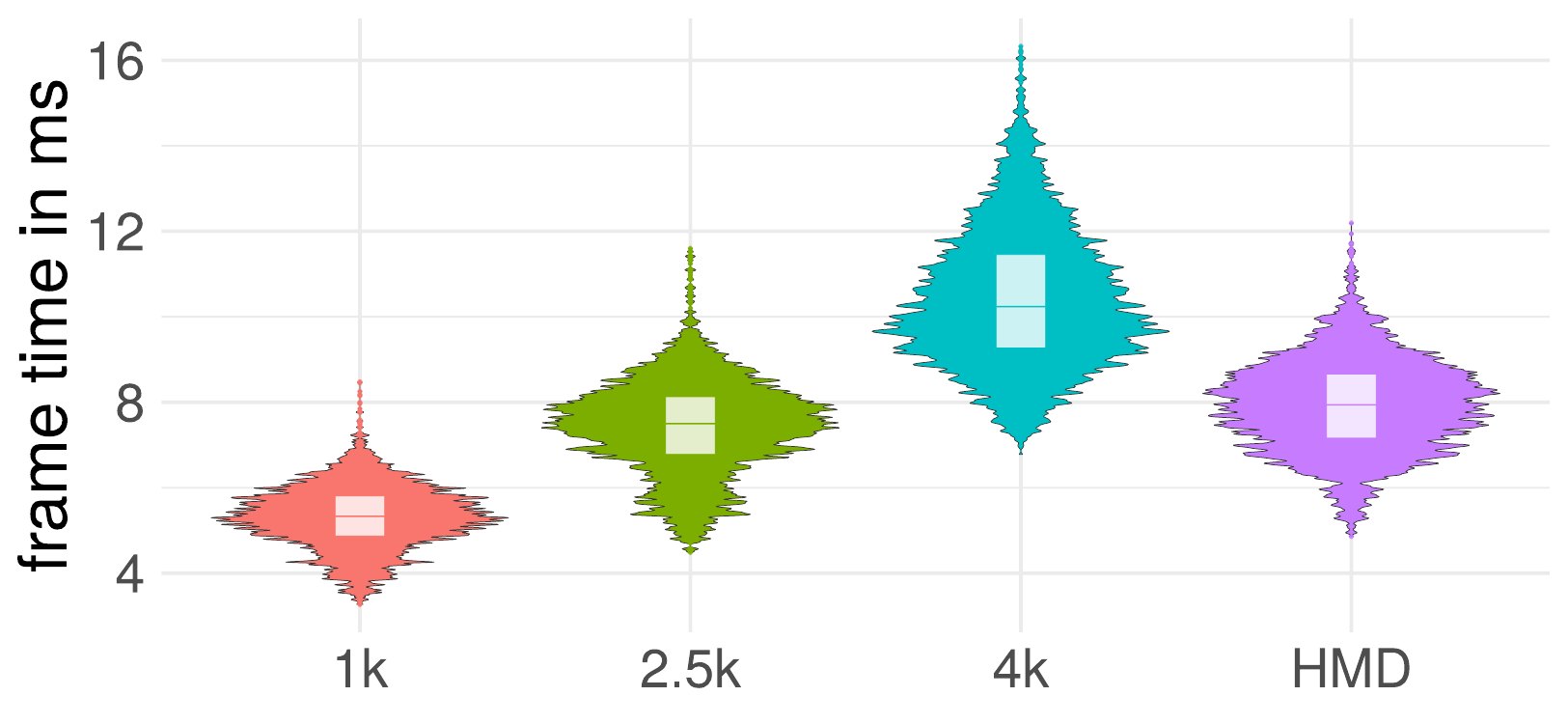} \\
                          &                    &  Triangles &  $656.20M$ &  $4.67M$ &  $5.63M$ & $10.02M$ &   $7.96M$ & \\
                          &                    &    Surfels &        $0$ &  $6.67M$ & $11.85M$ & $21.07M$ &   $7.60M$ & \\
                          &                    &        FPS &     $3.52$ & $152.90$ & $101.91$ &  $84.92$ &   $87.92$ & \\
                          &                    &    Quality &      $1.0$ &   $0.99$ &   $0.99$ &   $0.99$ &       $-$ & \\\hline
            
    \mr{4 - Bakery}       & \tblimg{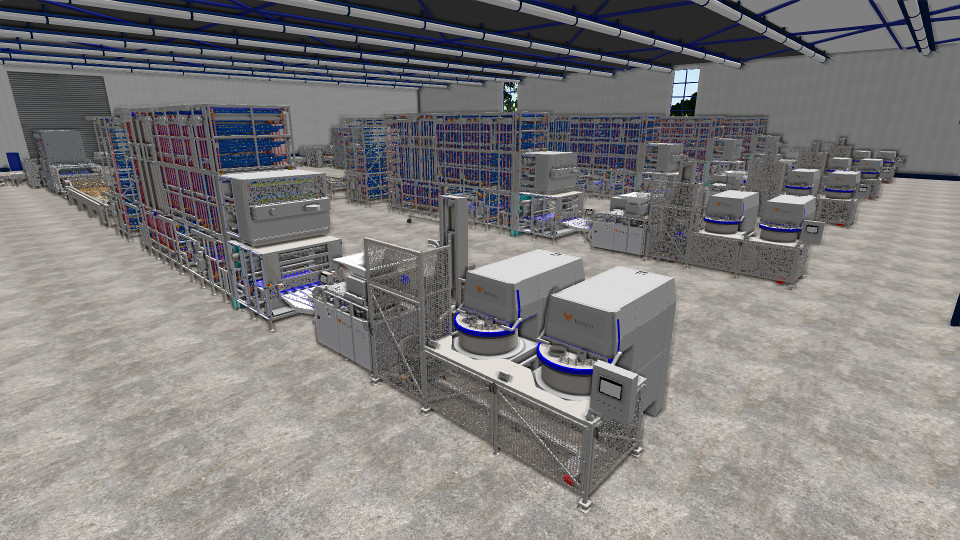} & Draw calls &    $51594$ &   $5693$ &   $8757$ &  $12781$ &   $10578$ & \tblimg{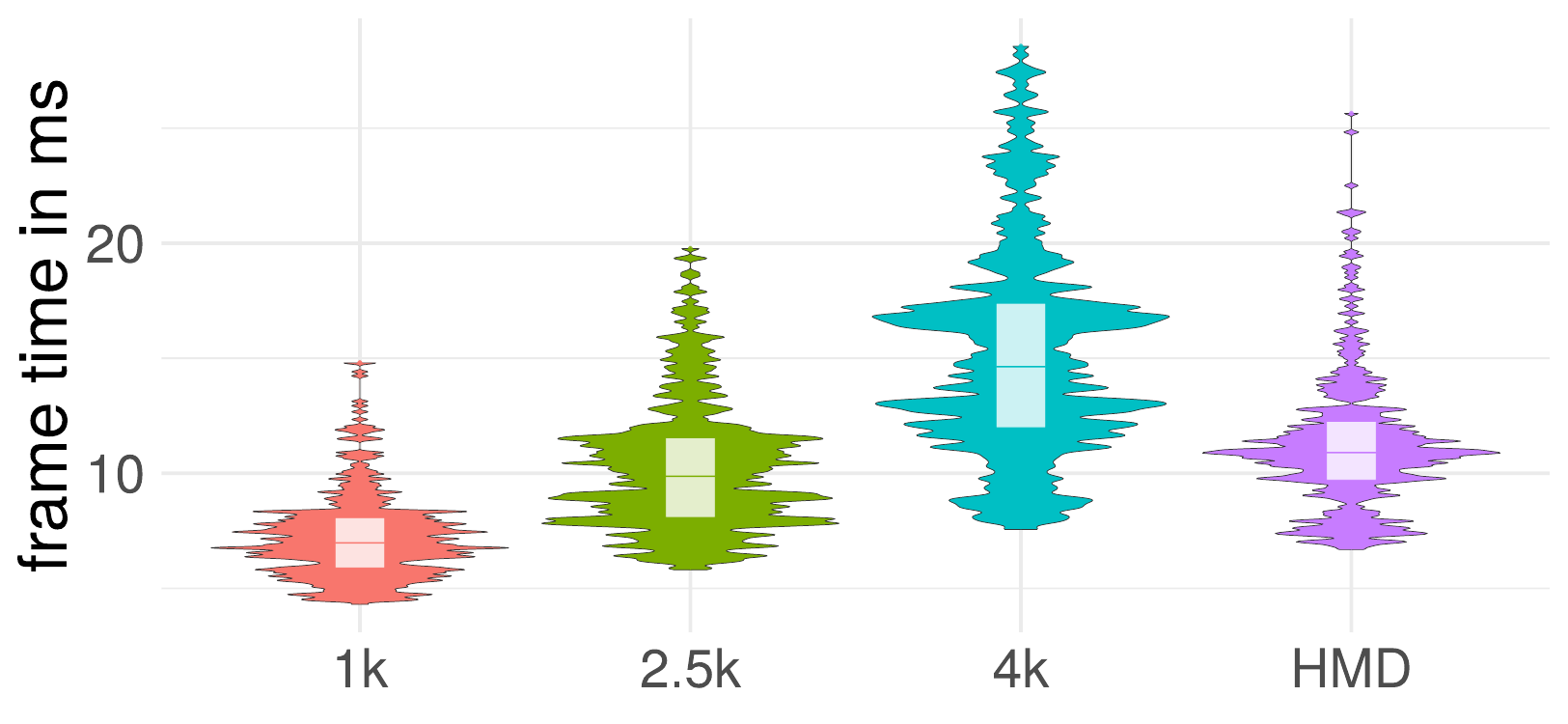} \\
                          &                    &  Triangles &  $949.09M$ & $11.18M$ & $18.17M$ & $31.21M$ &  $21.31M$ & \\
                          &                    &    Surfels &        $0$ & $11.31M$ & $18.08M$ & $32.55M$ &  $15.70M$ & \\
                          &                    &        FPS &     $2.39$ &  $70.13$ &  $44.23$ &  $32.03$ &   $37.78$ & \\
                          &                    &    Quality &      $1.0$ &   $0.89$ &   $0.91$ &   $0.95$ &       $-$ & \\\hline
            
    \mr{5 - Power Plants} & \tblimg{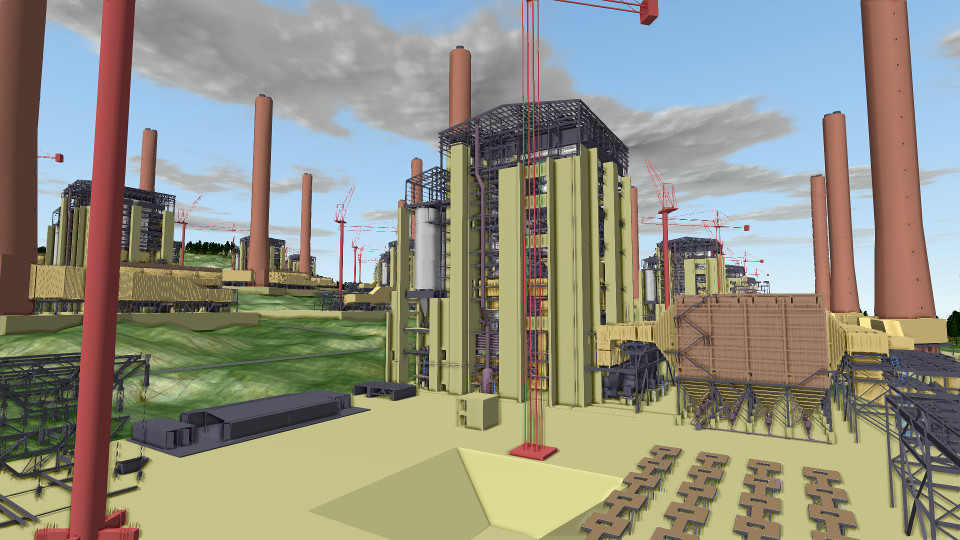} & Draw calls &     $8610$ &   $2360$ &   $4195$ &   $6477$ &    $3233$ & \tblimg{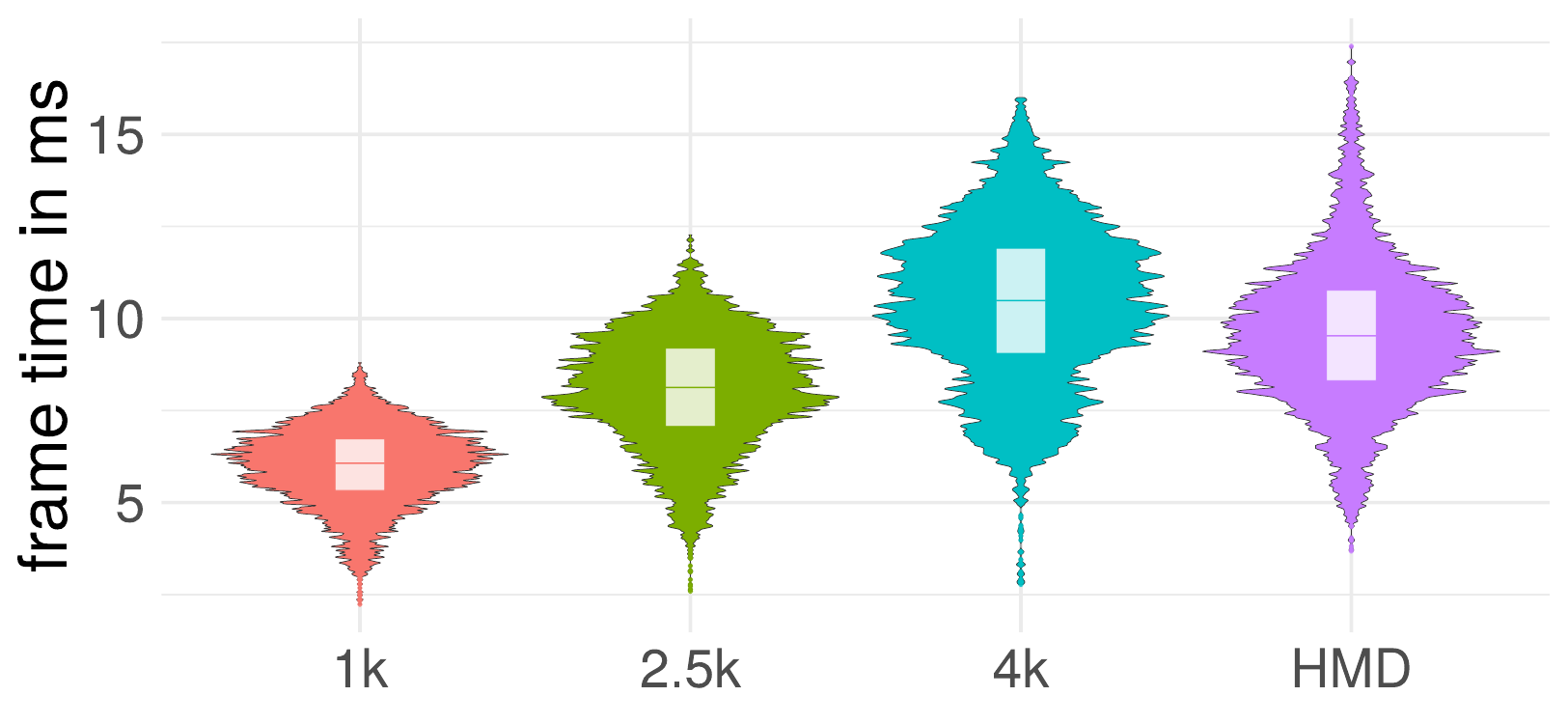} \\
                          &                    &  Triangles &  $419.44M$ &  $7.41M$ & $10.31M$ & $18.87M$ &  $13.40M$ & \\
                          &                    &    Surfels &        $0$ &  $6.49M$ &  $8.64M$ & $13.00M$ &  $10.11M$ & \\
                          &                    &        FPS &     $5.70$ & $150.24$ & $111.22$ &  $76.97$ &   $83.84$ & \\
                          &                    &    Quality &      $1.0$ &   $0.99$ &   $0.99$ &   $0.99$ &       $-$ & \\\hline
  \end{tabular}
\end{table*}

In this section we examine how our proposed method performs in terms of real-time rendering performance and image quality.
The image quality was measured by comparing the approximated image with the image rendered without LOD using the hierarchical Structural SIMilarity (SSIM) index method proposed by Wang et al. \cite{Wang2004}. 
An example of this method is shown in \autoref{fig:eval:quality}. 
The upper image shows a camera view rendered with our method, while the middle image shows a view without any approximations. 
The bottom image shows the SSIM image which highlights the differences of both images from which the SSIM index (0.81) is computed.
Especially noticable is the difference at the trees and the crane of the power plant.
The trees and crane rendered with our method seem more volumetric since our method cannot handle thin or finely detailed objects very well.
However, this fact can also be utilized to achieve some degree of anti-aliasing.

For the evaluation of our method, we placed cameras at various representative positions in our benchmark scene to show different aspects of our algorithm.
For each camera position we measured the average frame time to render a single image at different resolutions (1k, 2.5k, 4k) as well as stereo rendering for head-mounted displays ($2\by1344\by1600$).
We computed the SSIM index for these positions at each resolution, except for HMD rendering since the value might get a wrong impression due to the distortion of the HMD lenses.

\autoref{fig:eval:performance} shows the measured camera positions with the resulting statistics for number of draw calls, rendering time, and image quality of this specific view for each of the different resolutions.
The last column additionally shows a combined box \& violin plot of the distribution of frame times measured at multiple uniformly distributed positions at the specific region of the view for each of the cardinal directions.
For the overview, we measured ${\sim}10000$ positions at a height of 200m above ground with a slight downward tilt for each camera view.
For the other views, we measured ${\sim}1000$ positions (${\sim}100$ for the bakery) at ground level (2m).

\begin{figure}[ht]
	\centering
	\includegraphics[width=0.9\columnwidth]{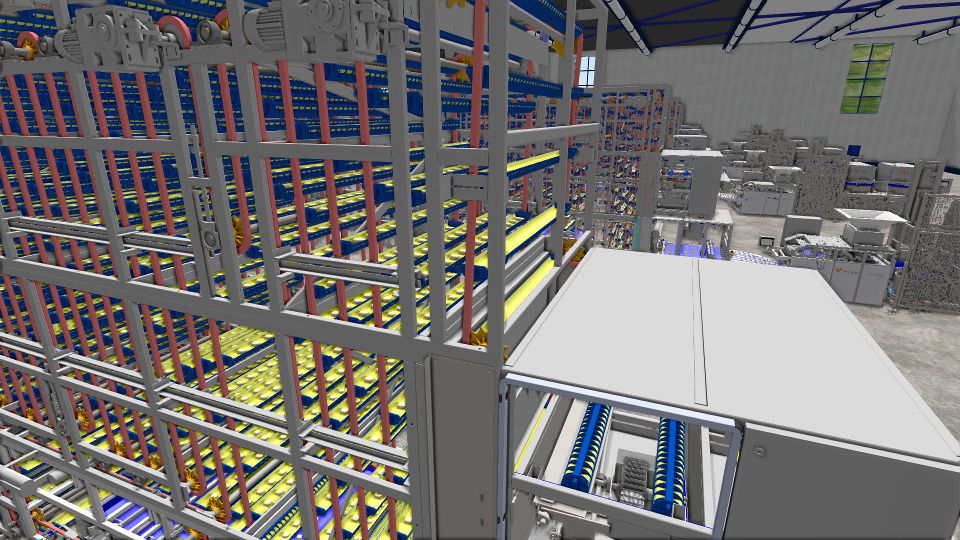}
	\caption{%
    Closeup of a donut production line from the \textit{Bakery} of our benchmark scene.
    Due to the high complexity in close vicinity to the observer, the scene part cannot be effectively approximated anymore by our algorithm.
  }
	\label{fig:eval:closeup}
\end{figure}

In general, we achieved high frame rates of at least 30 fps for each camera position and resolution while achieving a relatively good SSIM index of at least 0.68 for the overview (which is mainly due to the large amount of trees) and at least 0.9 for camera positions closer to the ground level.  
For resolution of $1920\by1080$ we even achieve frame rates of at least 60 fps throughout the entire scene.
The lowest frame rates can be observed at \textit{Pompeii} and the \textit{Bakery}, which are the most complex parts of our benchmark scene.
Here, we have high number of complex scene objects close to the observer, which cannot be effectively approximated anymore by our algorithm alone (see, e.g., \autoref{fig:eval:closeup}).
Still, the median frame times in these cases are still at least around 15 ms (${\sim}66\fps$) for a 4k resolution and at least 13 ms (${\sim}76\fps$) for HMD stereo rendering as can be seen in the box plot in the rightmost column of \autoref{fig:eval:performance}.

Although we did not achieve the desired 90 fps for stereo rendering on an HMD everywhere, we were still able too keep the frame rates (mostly) in the range of 45-90 fps which is the range used for effective time-warping to reduce nausea \cite{Waveren2016}.
These frame rates can certainly be improved using further specialized techniques for HMD rendering, like fixed foveated rendering or multiresolution framebuffers.

\section{Conclusion and future work}
\label{sec:conclusion}

We have presented an efficient point based algorithm for generating and rendering continuous approximations of highly complex scenes in real time, even for VR applications using head-mounted displays on standard consumer hardware.
Our method can handle a large variety of scene types, including complex CAD data.
It can robustly create approximations of almost any surface requiring only little user interaction and parameterization.
The image quality is reasonably good (with room for improvement) and, due to the continuous nature, there are almost no visible popping artifacts when navigating a scene.
Using our method combined with other culling and level-of-detail techniques, we are certain that we can achieve an even better performance, and with better point based filtering to improve the visual quality it might also be applicable in the context of games.

However, there are some limitations.
Since our method is mostly intended for objects with a small projected size on the screen, very complex geometry, which covers a large portion of the screen, cannot be effectively approximated.
Also, since we uniformly distribute points on the visible surface of an object, we might draw unnecessarily many points for long objects where one part is close to the observer while other parts are further away.
Currently, the only way to circumvent these problems is to cut these objectes into smaller parts and approximate each part seperately, which results in more draw calls and possibly higher memory consumption for the surfel approximations.

Another type of objects that cannot be handled well, are objects with thin surfaces or walls.
Since our sampling method currently does not incorporate the normal of a sampling point, it might happen that we incorectly distribute the surfels on the surface of such objects, which results in holes during rendering.
In future work, we want to include (approximate) geodetic distances in our generation of greedy permutations to better capture the surface of an object.

Furthermore, we want to include better filtering of color values of the surfels, since our current sampling process does not take the colors into account.
It can happen that samples are taken during the preprocessing step from pixels with colors of minor importance, as can be seen in \autoref{fig:render:prefixes} (blue disc in the first few pictures).
This could be circumvented, e.g., by averaging the color values of early prefixes in an additional post-processing step.
Another idea would be, to generate the samples from different mip-levels of the rendered images. 

Due to the progressive nature of our method, it is an ideal basis for streaming applications in out-of-core systems and mobile rendering.
In future work, we also want to use our method for on-the-fly generation of approximations at run time, e.g. for very large, procedurally generated worlds.

\bibliography{BlueSurfels}

\end{document}